\documentclass{article}

\usepackage{arxiv}

\usepackage[utf8]{inputenc} 
\usepackage[T1]{fontenc}    
\usepackage{hyperref}       
\usepackage{url}            
\usepackage{booktabs}       
\usepackage{amsfonts}       
\usepackage{nicefrac}       
\usepackage{microtype}      
\usepackage{lipsum}		
\usepackage{graphicx}
\usepackage{doi}
\usepackage{bm}
\usepackage{amssymb}
\usepackage{float}
\usepackage{amsmath}
\usepackage{multirow}

\usepackage{physics}
\usepackage{algorithm}
\DeclareMathOperator{\Var}{Var}

\usepackage{algpseudocode}  

\usepackage{multirow}
\usepackage[table,xcdraw]{xcolor}

\usepackage[table,xcdraw]{xcolor}

\title{An Uncertainty-Aware Physics-Informed Neural Network Solution for the Black–Scholes Equation: A Novel Framework for Option Pricing}

\date{April, 2025}	

\author{ \href{}{\hspace{1mm} Sina Kazemian\textsuperscript{*1}, Ghazal Farhani\textsuperscript{*2}, Amirhessam Yazdi\textsuperscript{3}}\\
 \textsuperscript{1}Department of Physics and Astronomy, University of Waterloo, Western University\\
  \textsuperscript{2} National Research Council Canada, London Ontario\\
  \textsuperscript{3} University of Nevada, Reno\\
}

\renewcommand{\shorttitle}{PINN's as an Alternative to the BS Model}

\begin{document}
\maketitle
\begingroup
  \renewcommand\thefootnote{\fnsymbol{footnote}}
  \footnotetext[1]{ \href{Leading Authors}{First two authors contributed equally to this paper}}
\endgroup
\footnotetext[1]{
\href{mailto:Sina Kazemian}{skazemi5@uwo.ca}}
\footnotetext[2]{
\href{mailto:Ghazal Farhani}{ghazal.farhani@nrc-cnrc.gc.ca}}
\footnotetext[3]{
\href{mailto:Amir Hessam Yazdi}{ayazdi@unr.edu}}

\begin{abstract}

We present an uncertainty-aware, physics-informed neural network (PINN) for option pricing that solves the Black--Scholes (BS) partial differential equation (PDE) as a mesh-free, global surrogate over $(S,t)$. The model embeds the BS operator and boundary/terminal conditions in a residual-based objective and requires no labeled prices. For American options, we handle early exercise via an obstacle-style relaxation while retaining the BS residual in the continuation region. To quantify \emph{epistemic} uncertainty, we introduce an anchored-ensemble fine-tuning stage (AT–PINN) that regularizes each model toward a sampled anchor, producing prediction bands alongside point estimates.

On European calls/puts, the approach achieves low errors (e.g., MAE $\sim$ $5{\times}10^{-2}$, RMSE $\sim$ $7{\times}10^{-2}$, explained variance $\approx 0.999$ in representative settings) and tracks ground truth closely across strikes and maturities. For American puts, the method remains accurate (MAE/RMSE on the order of $10^{-1}$ with EV $\approx 0.999$) and avoids the error accumulation typical of time-marching schemes. Against data-driven baselines (ANN, RNN) and a Kolmogorov--Arnold FINN variant (KAN), our PINN matches or outperforms on accuracy while training more stably; anchored ensembles provide uncertainty bands that align with observed error scales. We discuss design choices (loss balancing, sampling near the payoff kink), limitations, and extensions to higher-dimensional BS settings and alternative dynamics.

\end{abstract}

\maketitle


\section{\label{sec:level1}Introduction}

Option pricing is a central topic in financial mathematics, essential for hedging risk and determining the fair value of derivative instruments. Over the years, several models have been developed, including the Binomial Option Pricing Model \cite{CoxRossRubinstein1979,MuroiSuda2022,Bjork2020}, Monte Carlo simulation methods \cite{Boyle1977,Giles2015}, the Heston Model \cite{Heston1993,Gatheral2012,KellerRessel2011}, and Merton’s Jump-Diffusion Model \cite{Merton1976Jump,ContTankov2004,Kou2002}. Among these, the Black–Scholes (BS) formula \cite{BlackScholes1973} remains the most widely adopted and has become the cornerstone of option pricing \cite{Bjork2020,Hull2022,Kumiega2024}. Its enduring influence stems from analytical tractability, computational efficiency, and ease of implementation, particularly for European-style options.

Despite its effectiveness, the BS model is constrained by its simplifying assumptions and its classical closed-form solutions. It does not naturally accommodate early exercise, path dependence, or various market frictions and nonlinearities (e.g., transaction costs, stochastic volatility, and jumps). American options, for example, allow the holder to exercise at any time before expiration, a feature that significantly increases modeling complexity. To handle such cases, numerical methods such as binomial trees and finite-difference schemes are widely used\cite{LeisenReimer1996,TavellaRandall2000,Duffy2006}, but they often face challenges related to convergence, stability, and computational cost \cite{PooleyVetzalForsyth2003,ClarkeParrott1999,ForsythVetzalZvan2002}. In addition, for American-style features in high dimensions, Monte Carlo methods with regression—most notably the Least-Squares Monte Carlo (LSMC) approach of Longstaff and Schwartz—are widely used \cite{LongstaffSchwartz2001}.

To address these limitations, Physics-Informed Neural Networks (PINNs) \cite{Raissi2019,Karniadakis2021,farhani2025simple,farhani2022momentum,bai2022application,alonso2023physics,hainaut2024option}
have emerged as a promising alternative. By embedding governing equations directly into the training objective, PINNs enforce consistency with financial PDE structure through residual and boundary/terminal penalties while leveraging the flexibility of deep neural networks. This hybrid formulation can improve data efficiency and deliver accurate surrogates in settings where closed-form solutions are unavailable or meshing is cumbersome (e.g., irregular domains, parameter sweeps, or higher dimensions). PINNs naturally extend to nonlinear PDEs, can reduce computational effort in high-dimensional problems, and adapt to diverse financial instruments through customizable architectures and optimization strategies. Beyond finance, physics-driven modeling across domains demonstrates the value of respecting governing operators in predictive modeling (e.g., PDE and integro–differential formulations in materials and condensed matter) \cite{kazemian2024influence,kazemian2023dynamic,kazemian2019thermal}. This perspective motivates embedding the BS operator within a learning framework.

However, a critical challenge with deep learning models, including PINNs, is their tendency toward overconfidence in predictions \cite{Guo2017,guynn2015google,Ovadia2019,gal2016uncertainty}. Conventional evaluation metrics such as root mean squared error or accuracy do not capture the reliability of outputs. In high-stakes domains like finance, quantifying predictive confidence is as important as accuracy itself. This requires to distinguish \emph{epistemic} uncertainty (from limited data or model misspecification) from \emph{aleatoric} uncertainty (inherent noise) \cite{KendallGal2017}. Bayesian neural networks provide a principled solution by treating parameters as probability distributions \cite{mackay1992practical,farhani2023bayesian}, but their practical use in large-scale networks is computationally prohibitive. Consequently, approximate methods, including ensembling, Monte Carlo dropout, and anchored ensembling, have gained traction \cite{gal2016uncertainty, pearce2020uncertainty, tsuchida2018invariance}. Anchored ensembling is especially noteworthy as it leverages MAP estimation. Although MAP has been extensively applied in inverse problems \cite{farhani2019optimal, gelb1974applied, tarantola2005inverse}, its adoption in machine learning frameworks has been more limited.

Building on these ideas, our study develops a PINN framework for pricing European and American options, benchmarked against the BS formula. The PINN incorporates the BS PDE directly into its loss function, enabling faithful reproduction of European option values under idealized assumptions. For American options, where early exercise introduces free-boundary conditions, the PINN provides a flexible approximation framework unconstrained by discretization. To address uncertainty, we extend the model with anchored ensembles, allowing estimation of confidence intervals for option values alongside point predictions. This dual capability—accurate pricing and reliable uncertainty quantification—offers a novel contribution to option pricing.

The remainder of this paper is organized as follows. Section~\ref{sec:related_work} reviews the previous reserach work on implementing deep leanring models for BS option pricing. Then, in Section~\ref{sec:preliminary}  theoretical foundations of the BS model and PINNs. Section 3 details the methodology, including optimization strategies and network architectures. Section 4 presents numerical experiments comparing PINNs with classical methods across multiple scenarios. Section 5 concludes with insights and future research directions.

\section{Related Work} \label{sec:related_work}

Recent work on data‐driven option pricing commonly adopts an artificial neural network ANN baseline that mirrors the BS input set \cite{gross2025comparative,rigopoulos2024ann,shvimer2024pricing,elbayed2025deep,d2025predicting}. For example, D’Uggento \textit{et al.} \cite{d2025predicting} define \textsc{ANN1} as a feedforward network trained on precisely the BS covariates—underlying price, strike, time to maturity, annual volatility, and a call/put flag—while deliberately excluding dividends and broader firm/market features so that any incremental value of non-BS information can be assessed against a clean benchmark. Such baselines are typically evaluated with point-prediction metrics (MAE, MSE, RMSE, $R^2$, MAPE) and seldom report calibrated predictive intervals or other forms of uncertainty quantification.

A \emph{recurrent} approach treats pricing as a sequence-learning problem: given per-date descriptors (e.g., moneyness, time to maturity, volatility proxies, and the call/put indicator), a recurrent neural network RNN with LSTM/GRU cells maintains a hidden state to capture temporal dependence and forecast current prices or implied volatilities \cite{pimentel2025option,liu2023option,zhang2021option,vinje2025merged}.On large empirical benchmarks, RNNs often outperform static ANNs and, in purely predictive terms, can surpass Black--Scholes on error metrics—reflecting benefits from modeling temporal dynamics and market frictions \cite{d2025predicting}. These gains come with familiar caveats: (i) sensitivity to rolling-window design and leakage control; (ii) absence of no-arbitrage and shape constraints (e.g., call-price monotonicity/convexity in strike, calendar monotonicity) unless explicitly regularized; (iii) higher computational cost; and (iv) continued reliance on point estimates without calibrated uncertainty.

Beyond purely data-driven models, physics- and finance-informed networks inject structure into learning. Dhiman \textit{et al.} \cite{dhiman2023physics} applied \emph{PINNs} to the BS equation; while results were promising, systematic benchmarking against alternative learning baselines and American-style early-exercise settings was limited. Finance-Informed Neural Networks (FINNs) \cite{aboussalah2024ai} push further by replacing a pure PDE-residual objective with a finance-aware loss that encodes no-arbitrage and self-financing hedging: a pricing network $g_\theta$ yields Greeks via automatic differentiation, and training penalizes hedging residuals of zero-cost $\Delta$ or $\Delta\Gamma$ portfolios across time. This yields a hybrid, self-supervised objective that leverages data while hard-wiring financial structure, demonstrated on synthetic European-option datasets under GBM and Heston dynamics. Training efficiency and stability, however, can be sensitive to architectural choices and hedging-instrument design, and—as with most deep-learning studies—uncertainty quantification is rarely reported.

To improve trainability within a structured paradigm, Liu \textit{et al.} \cite{liu2024kolmogorov} introduce a Kolmogorov–Arnold Finance-Informed Neural Network(\emph{KAFIN}), which parameterizes the pricing map using the Kolmogorov–Arnold representation: the multivariate function is expressed as sums of learned univariate transforms composed with learned univariate aggregators. Practically, this Kolmogorov–Arnold Network (KAN) architecture separates \emph{inner} per-input scalar transforms from \emph{outer} scalar mixers and is trained with a composite, PINN-style objective that can include initial/terminal and boundary conditions together with a finance-governed operator term (e.g., the BS operator) and, optionally, FINN-type hedging residuals \cite{liu2024kolmogorov}. On controlled, simulated European-option data, KAFIN reports lower losses and competitive or improved training efficiency relative to baseline FINN settings, at the cost of extra architectural hyperparameters and tuning sensitivity. While some variants discuss extensions, systematic evaluation on live market data, American early-exercise features, and calibrated uncertainty remains comparatively limited \cite{aboussalah2024ai,liu2024kolmogorov}.

\emph{Our approach} follows the PINN line \cite{dhiman2023physics,aboussalah2024ai,liu2024kolmogorov} but adopts a strictly physics-informed formulation that directly solves the BS PDE by minimizing a residual-based loss augmented with initial and boundary conditions. We parameterize a mesh-free surrogate $V_\theta(S,t)$ and, without additional hedging residuals or bespoke regularizers, accurately recover both European and American prices while enforcing arbitrage-consistent bounds through our output mapping. In contrast to purely data-driven ANN/RNN baselines and the hedging-based FINN family, our method does not require labeled prices or hedging targets and naturally supports standard model variants. Crucially, we move beyond point estimates by quantifying epistemic uncertainty, producing uncertainty bands around predictions; our anchor-augmented objective further stabilizes training and enables fast transfer fine-tuning.

\section{Preliminaries}\label{sec:preliminary}
We begin by recalling the BS pricing PDE and show how it is enforced in a physics-informed neural network through a residual-based loss. We then introduce an anchor-augmented objective for uncertainty quantification and efficient transfer.

\subsection{Black--Scholes model}

Under the risk–neutral measure, the underlying price \(S(t)\) follows a geometric Brownian motion on $t\in[0,T)$,
\begin{equation}
    dS(t)=r\,S(t)\,dt+\sigma\,S(t)\,dW(t),
\end{equation}
where $r$ is the risk–free rate and $\sigma>0$ the volatility. This yields the BS PDE for a European option price \(V(S,t)\):
\begin{equation}
    \frac{\partial V}{\partial t}+\frac{1}{2}\sigma^2 S^2 \frac{\partial^2 V}{\partial S^2}
    + r S \frac{\partial V}{\partial S} - rV = 0,
    \label{equ:BS_PDE_Model}
\end{equation}
with terminal condition \(V(S,T)=\max(S-K,0)\) for a call. The BS framework admits closed forms for European options, while early–exercise features (e.g., American options) lead to free–boundary problems that typically require numerical methods.

\subsection{PINNs for option pricing}

Let \(\hat V_\theta(S,t)\) be a neural approximation of \(V(S,t)\) with trainable parameters \(\theta\).
Define the BS residual operator
\begin{equation}
    \mathcal{R}_{\text{BS}}[\hat V_\theta](S,t)
    \;=\;
    \frac{\partial \hat V_\theta}{\partial t}
    + \frac{1}{2}\sigma^2 S^2 \frac{\partial^2 \hat V_\theta}{\partial S^2}
    + r S \frac{\partial \hat V_\theta}{\partial S} - r\,\hat V_\theta .
\end{equation}

Given collocation points \(\{(S_i,t_i)\}_{i=1}^{N_r}\) in the interior, terminal samples \(\{(S_i,T)\}_{i=1}^{N_i}\), and boundary samples \(\{(S_b,t_b)\}_{b=1}^{N_b}\) with targets \(h(S_b,t_b)\) (Dirichlet or Neumann), the standard PINN objective is
\begin{equation}
\begin{split}
\label{loss_PDE}
\mathcal{L}_{\text{PINN}}(\theta)
=
& \lambda_r\,\frac{1}{N_r}\sum_{i=1}^{N_r}\!\bigl(\mathcal{R}_{\text{BS}}[\hat V_\theta](S_i,t_i)\bigr)^2
+
\\ &\lambda_i\,\frac{1}{N_i}\sum_{i=1}^{N_i}\!\bigl(\hat V_\theta(S_i,T)-\max(S_i-K,0)\bigr)^2
+
\lambda_b\,\frac{1}{N_b}\sum_{b=1}^{N_b}\!\bigl(\hat V_\theta(S_b,t_b)-h(S_b,t_b)\bigr)^2,
\end{split}
\end{equation}
where $N_r,N_i,N_b$ are the counts of interior, terminal, and boundary samples, respectively, and $\lambda_r,\ \lambda_i,$ and $\lambda_b$ are positive weighting coefficients that balance the relative contributions of the PDE‐residual, terminal, and boundary losses in $\mathcal{L}_{\text{PINN}}$ (i.e., they control each term’s influence during training). This mesh–free loss enforces the PDE, terminal payoff, and boundary conditions via automatic differentiation.

\subsection{Uncertainty quantification via anchoring}

Anchored ensembling \cite{pearce2020uncertainty} approximates Bayesian inference by adding a quadratic pull toward a reference parameter vector (an ``anchor''), producing MAP-like estimators whose ensemble spread reflects posterior uncertainty. For regression, a typical anchored loss for the \(j\)-th model is
\begin{equation}
    \mathcal{L}_{\text{anch},j}(\theta_j)
    =
    \frac{1}{N}\sum_{n=1}^{N}\bigl(y_n-\hat y_{\theta_j}(x_n)\bigr)^2
    + \frac{\lambda_{\text{anc}}}{N_\theta}\,\lVert \theta_j-\theta_{\text{anc},j}\rVert_2^2,
\end{equation}
where \(\theta_{\text{anc},j}\) is drawn from a prior (or otherwise specified). In our PINN setting we combine this idea with the residual/condition losses:
\begin{equation}
\label{equ:anchor_PINN_Loss}
\mathcal{L}_{\text{AT-PINN}}(\theta)
=
\mathcal{L}_{\text{PINN}}(\theta)
+\frac{\lambda_{\text{anc}}}{N_\theta}\,\lVert \theta-\theta_{\text{anc}}\rVert_2^2,
\end{equation}
where \(\theta_{\text{anc}}\) is a chosen anchor (e.g., a data–dependent anchor from a first training stage; see Sec.~\ref{sec:method}). Setting \(\lambda_{\text{anc}}{=}0\) recovers \eqref{loss_PDE}.

\section{Methodology} \label{sec:method}

\subsection{Problem setup}
We approximate \(V(S,t)\) for European and American options with a fully connected network \(\hat V_\theta\). Training samples comprise interior collocation points for the PDE, terminal points at \(t{=}T\) for the payoff, and boundary points at \(S{=}S_{\min},S_{\max}\) for boundary conditions. Unless otherwise noted, we work with the formulation of \((\log S,t)\).

\subsection{Two–stage training with anchoring transfer} \label{sec:train}
To obtain fast convergence and calibrated uncertainty, we adopt a two–stage procedure summarized in Alg.~\ref{alg:atpinn}.
\begin{itemize}
    \item We train \(\hat V_\theta\) by minimizing \(\mathcal{L}_{\text{PINN}}(\theta)\) of Eq~\ref{loss_PDE} and save the converged parameters \(\theta^{(1)}\).
    \item We instantiate a second model, initialize \(\theta\leftarrow\theta^{(1)}\), set the anchor \(\theta_{\text{anc}}:=\theta^{(1)}\), and fine–tune with the anchor–augmented loss \(\mathcal{L}_{\text{AT-PINN}}(\theta)\) of Eq~\ref{equ:anchor_PINN_Loss}. The transfer initialization provides a strong prior, so Stage~2 requires substantially fewer epochs (e.g., $5{,}000$ vs.\ $50{,}000$).
    \item We repeat Stage~2, \(M\) times (resampling batches and seeds) to obtain predictors \(\{\hat V_{\theta^{(m)}}\}_{m=1}^M\); we report the ensemble mean and standard deviation as point estimate and predictive uncertainty.
\end{itemize}

\begin{algorithm}[h]
\caption{Anchor–Transfer PINN (AT–PINN) for Black--Scholes}
\label{alg:atpinn}
\begin{algorithmic}[1]
\Require BS params $(r,\sigma)$, strike $K$; domains for $(S,t)$; data sets $\mathcal{D}_r,\mathcal{D}_i,\mathcal{D}_b$; weights $\lambda_r,\lambda_i,\lambda_b,\lambda_{\text{anc}}$; ensemble size $M$; optimizer $\mathcal{O}$
\Ensure Ensemble mean $\mu(S,t)$ and std $\varsigma(S,t)$ of $\hat V(S,t)$
\State Define residual $\mathcal{R}_{\text{BS}}[\hat V_\theta]$ and loss $\mathcal{L}_{\text{PINN}}(\theta)$ as in \eqref{loss_PDE}
\State \textbf{Stage 1:} initialize $\theta\leftarrow\theta^{(0)}$; for $E_1$ epochs update $\theta\leftarrow\mathcal{O}\!\bigl(\theta,\nabla_\theta \mathcal{L}_{\text{PINN}}(\theta)\bigr)$
\State Save $\theta^{(1)}\leftarrow\theta$ and set anchor $\theta_{\text{anc}}\leftarrow\theta^{(1)}$
\State \textbf{Stage 2:} \For{$m=1$ to $M$}
  \State \quad Initialize $\theta\leftarrow\theta^{(1)}$
  \State \quad For $E_2$ epochs minimize $\mathcal{L}_{\text{AT-PINN}}(\theta)=\mathcal{L}_{\text{PINN}}(\theta)+\frac{\lambda_{\text{anc}}}{N_\theta}\lVert\theta-\theta_{\text{anc}}\rVert_2^2$
  \State \quad Store predictor $\hat V^{(m)}(S,t)\leftarrow \hat V_\theta(S,t)$
\EndFor
\State Aggregate $\mu(S,t)=\frac{1}{M}\sum_{m=1}^M \hat V^{(m)}(S,t)$ and $\varsigma(S,t)=\sqrt{\frac{1}{M-1}\sum_{m=1}^M\bigl(\hat V^{(m)}(S,t)-\mu(S,t)\bigr)^2}$
\end{algorithmic}
\end{algorithm}

\subsection{Neural architecture and sampling}
\begin{figure}[H]
    \centering
    \includegraphics[width=0.55\linewidth]{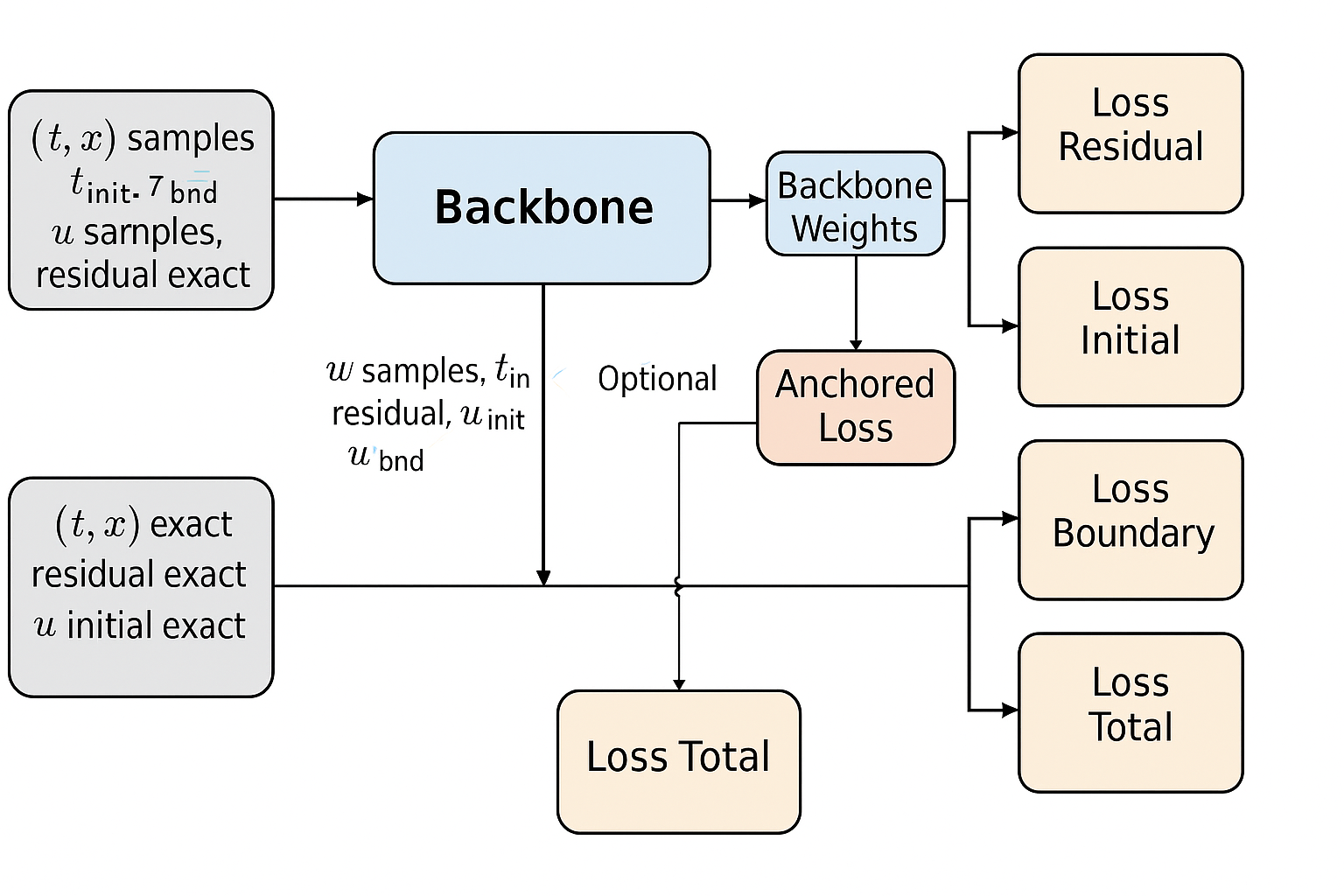}
    \caption{Schematic of the physics-informed neural network used to approximate $V(S,t)$: 
    inputs $S,t$, four \texttt{tanh} hidden layers (50 units each), and a single scalar output $\hat V_\theta$.}
    \label{fig:PINN_blocks}
\end{figure}

We use a fully connected feedforward network with four hidden layers (50 units each, \texttt{tanh} activations). 
The network takes $(S,t)$ as inputs and returns a single scalar output $\hat V_\theta(S,t)$. 
We train with Adam (initial learning rate $10^{-3}$, exponential decay), using $\lambda_r,\lambda_i,\lambda_b$ in Eq.~\ref{loss_PDE} and $\lambda_{\text{anc}}$ in Eq.~\ref{equ:anchor_PINN_Loss} as specified per experiment. 
Training sets comprise interior collocation points $N_r$, terminal samples $N_i$ at $t=T$, and boundary samples $N_b$ at $S_{\min},S_{\max}$. 
For American options (Sec.~\ref{sec:Amer-deter}), we additionally enforce $V\ge \text{payoff}$ via a projection step while retaining the PDE residual in the continuation region. 
A schematic overview of this architecture is shown in Fig.~\ref{fig:PINN_blocks}.

\subsection{Metrics and Uncertainty}


Let $\{(x_i,y_i)\}_{i=1}^N$ be targets and $\{\hat y_i\}_{i=1}^N$ predictions. Then
$\mathrm{MAE}=\tfrac{1}{N}\sum_{i=1}^{N}\lvert y_i-\hat y_i\rvert$; 
$\mathrm{RMSE}=\sqrt{\tfrac{1}{N}\sum_{i=1}^{N}(y_i-\hat y_i)^2}$; 
$\mathrm{EV}=1-\dfrac{\operatorname{Var}(y-\hat y)}{\operatorname{Var}(y)}$; 
and the relative error (\%) $=\dfrac{100}{N}\sum_{i=1}^{N}\dfrac{\lvert y_i-\hat y_i\rvert}{\lvert y_i\rvert}$. 

For an ensemble of $M$ models with pointwise predictions $\{\hat y^{(m)}(x)\}_{m=1}^M$,
\begin{align}
\bar y(x) 
&= \frac{1}{M}\sum_{m=1}^{M}\hat y^{(m)}(x), \\[4pt]
\widehat{\sigma}^2(x) 
&= \frac{1}{M-1}\sum_{m=1}^{M}\Bigl(\hat y^{(m)}(x) - \bar y(x)\Bigr)^2.
\end{align}
We visualize uncertainty as $\bar y(x)\pm k\,\widehat{\sigma}(x)$ with $k\in\{1,2\}$.

For any set $\{z_i\}_{i=1}^N$, the (sample) variance is
\begin{equation}
\Var(z) \;=\; \frac{1}{N-1}\sum_{i=1}^{N}\bigl(z_i - \bar z\bigr)^2,
\qquad
\bar z \;=\; \frac{1}{N}\sum_{i=1}^{N} z_i.
\end{equation}

\section{Result}
All experiments were implemented in Python with TensorFlow, enforcing the PDE via automatic differentiation. Training ran on NVIDIA T4 GPUs (Google Colab). The codebase is modular—separating PDE operators, boundary/initial conditions, and the training loop—so it can be readily adapted to other pricing PDEs or model variants (GitHub link).

In Sections~\ref{sec:Euro-deter} and \ref{sec:Amer-deter}, we report results using the \emph{anchored} PINN loss of Eq.~\ref{equ:anchor_PINN_Loss}, which is key to our contribution: beyond accurate point estimates, the model provides \emph{epistemic uncertainty quantification}. Concretely, predictions are returned with calibrated uncertainty bands that (i) respect no-arbitrage bounds through our bounded-output mapping and (ii) reflect model uncertainty arising from model behaviour. These intervals make the outputs actionable—supporting risk-aware decisions and enabling coverage/consistency checks—rather than merely offering single-value forecasts.

\subsection{European Option} \label{sec:Euro-deter}

\begin{figure}[h!]
    \centering
    \includegraphics[width=0.8\linewidth]{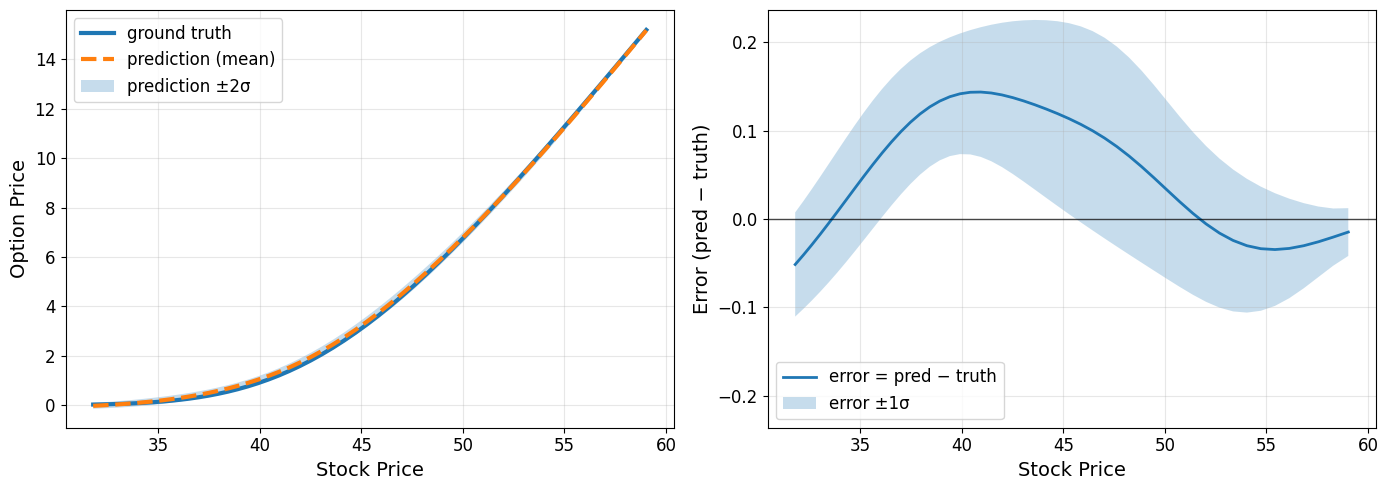}

    \caption{Left Panel: Call price vs stock price shown for ground truth in blue and the prediction in orange curve, with the shaded light blue representing uncertainty at $pm2$ for better visualization. Right Panel: Absolute error $(\hat{y} - y)$ at each stock price between the ground truth and the prediction is plotted. }
    \label{fig:european_option}
\end{figure}

We price a European put with six months to expiry, volatility $\sigma=0.2$, risk-free rate $r=0.05$, and strike $K=\$45$. The PINN is trained with $150$ interior (collocation) points and $256$ boundary/terminal points. Figure~\ref{fig:european_option} (top) compares the PINN prediction (orange) with the ground-truth solution (blue). We also construct an ensemble of $M=30$ independently trained models using an anchored loss; the light-blue band shows the $\pm 2\sigma$ ensemble uncertainty around the mean prediction $\bar{y}$. Because the uncertainty is only a few cents, we display $\pm 2\sigma$ for visibility.

The bottom panel plots the signed error $e_i=\hat{y}_i-y_i$ and overlays a $\pm 1\sigma$ band to indicate predictive uncertainty at each $x_i$. The uncertainty is computed on the predictions; for visualization, we show it around both the prediction and the error.

Across the slice, errors are on the order of a few cents with the relative percentage error not exceeding 1.2\%. Moreover, we report MAE = $5.21\times 10^{-2}$, RMSE = $7.12\times 10^{-2}$ and explained variance $\mathrm{EV}=0.999$.

\subsection{American Option }\label{sec:Amer-deter}
\begin{figure}
    \centering
    \includegraphics[width=1\linewidth]{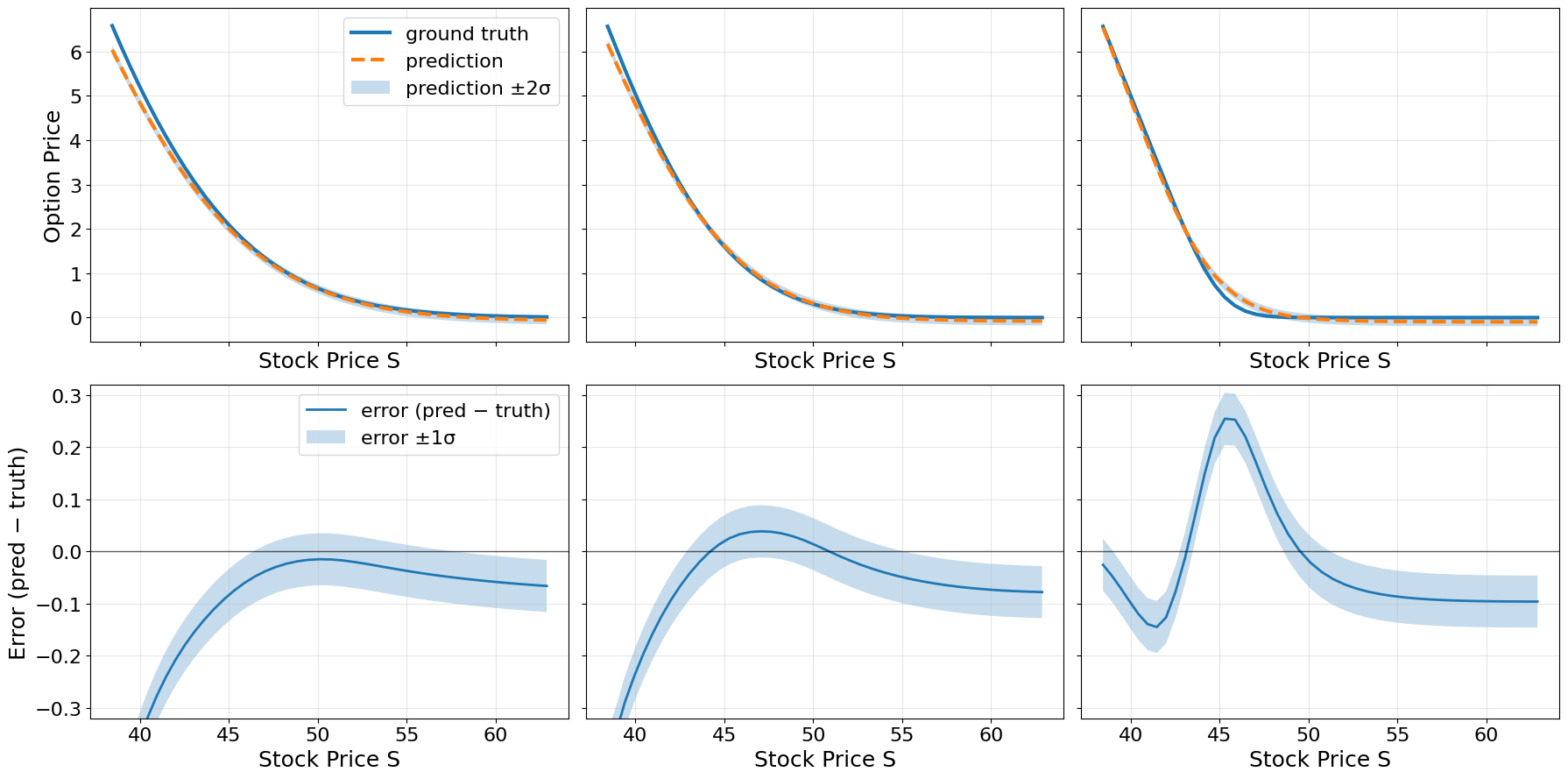}
    \caption{Top Panel: Orange dashed line is the prediction, and the solid blue line is the groundtruth with a shaded model uncertainty shown at 2 standard deviations. Bottom Panel: The error between the ground truth and the prediction and the shaded region is the uncertainty of the prediction at 1 standard deviation. In both panels, the left figure shows $t = 0$, middle panel shows midway to the expiry date, and the right panel is the expiry date.}
    \label{fig:American_Put}
\end{figure}

For the American-option study, we analyze a \emph{put} to emphasize its defining difference from the European case—namely, the right to exercise at any time before expiry. We adopt the same market and training settings as in the European benchmark: maturity \(T=0.5\) years, volatility \(\sigma=0.2\), risk-free rate \(r=0.05\), and strike \(K=\$45\). The numbers of collocation points, boundary-condition points, and training epochs are kept identical. Figure~\ref{fig:American_Put} (top) compares the PINN solution (orange) with a finite-difference reference (blue) at three time slices: \(t=0\) (present), \(t=T/2\), and \(t=T\) (maturity). The blue shading denotes predictive uncertainty; the corresponding errors are shown in the bottom panel.

At \(t=0\), the PINN slightly underprices for \(S<K\) (on average by \(\sim\$0.10\)), though overall errors remain small. At \(t=T/2\), predictions are generally below the reference, with a mild overvaluation near the exercise kink at \(S\!\approx\!K\) (under \(\$0.10\)). At maturity \(t=T\), the local overvaluation near \(S\!\approx\!K\) increases (up to \(\$0.30\)), a slice that is of limited practical relevance. Unlike time-marching numerical schemes—where backward stepping can accumulate error—the PINN shows no systematic degradation away from maturity. Overall performance is strong: \(\mathrm{MAE}=8.05\times10^{-2}\), \(\mathrm{RMSE}=9.11\times10^{-2}\), and explained variance \(\mathrm{EV}=0.999\).

\subsection{Comparison with Other Methods}
To assess our approach, we compare against three baselines: the Kolmogorov–Arnold FINN of \cite{liu2024kolmogorov} (henceforth \textbf{KAN}), a simple feed-forward network (\textbf{ANN}), and a recurrent model (\textbf{RNN}) following \cite{d2025predicting,gross2025comparative}. We build supervised datasets for American \emph{puts} and European \emph{calls} on the Yahoo Finance tickers \texttt{AAPL, MSFT, GOOGL, AMZN, META, F, T, NOK, CCL, SOFI, SNAP, NIO, AAL, PBR, RIVN}. 

\begin{figure}[h]
    \centering
    \includegraphics[width=0.46\linewidth]{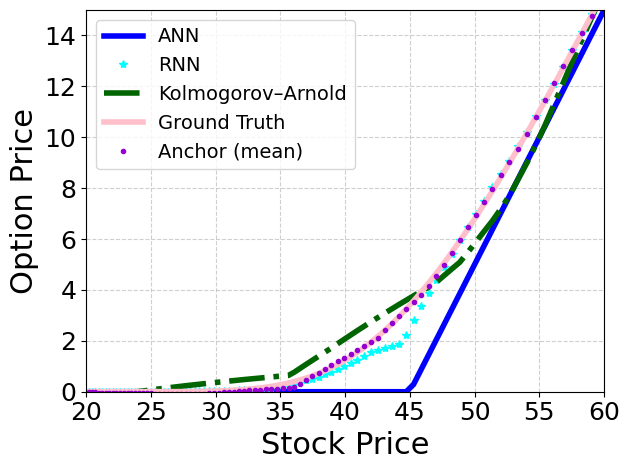}
    \includegraphics[width=0.45\linewidth]{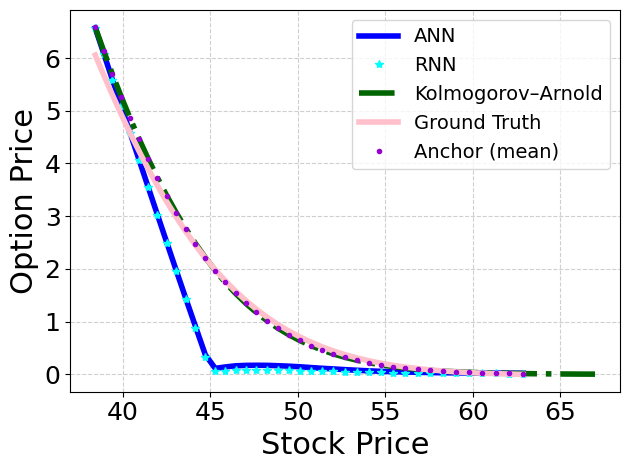}
    \caption{Ground truth (pink), \textbf{KAN} (dashed dark green), \textbf{PINN-anchor} (dotted purple), \textbf{RNN} (cyan stars), and \textbf{ANN} (solid blue). Left Panel: European, Right Panel: American}
    \label{fig:comparison_models}
\end{figure}

For each expiry–strike pair we extract the relevant option chain (put for American, call for European) and set the label to the mid-quote \((\text{bid}+\text{ask})/2\), falling back to the last traded price when necessary. To enforce no-arbitrage under a non-dividend assumption, labels \(y\) are clipped to instrument-specific bounds—American put: \(L=\max(K-S,0)\), \(U=K\); European call: \(L=\max(S-K e^{-r\tau},0)\), \(U=S\). Each sample includes static covariates \([\log(S/K),\,\tau,\,\sigma]\) (spot, time to maturity in years, implied volatility). For stable training with guaranteed bounded predictions, we map prices via \(f=(y-L)/(U-L)\in(0,1)\), \(z=\log\!\frac{f}{1-f}\in\mathbb{R}\), train on \(z\), and invert at inference by \(\hat y=L+(U-L)\,\sigma(z)\). For the RNN branch we augment each row with a per-ticker sequence of daily log-returns and realized volatility (rolling standard deviation, annualized by \(\sqrt{252}\)) using the last \(W\) days (e.g., \(W=60\)); sequences are standardized with train-split moments, and a fallback is used if a ticker’s history is missing. Across all experiments we fix the neural architecture per model and do not retune it per parameter set.

Model architectures are aligned to ensure a fair comparison. The \textbf{ANN} flattens the \((w,f_{\text{seq}})\) window (temporal order ignored) and processes it with Dense(128, ReLU) \(\rightarrow\) Dense(64, ReLU); the static branch is Dense(32, ReLU). The embeddings are concatenated and passed to a fusion head Dense(64, ReLU) \(\rightarrow\) Dense(32, ReLU) \(\rightarrow\) Dense(1) (linear) that predicts the scalar target \(z\). The \textbf{RNN} replaces the flattening with \(\mathrm{LSTM}(64)\) to capture temporal dynamics before the same fusion head; Dense layers use ReLU and the LSTM retains its standard tanh/sigmoid internals. Our \textbf{KAN} follows the Kolmogorov–Arnold form \(u(\mathbf{x}) \approx \sum_{i=1}^{M} g_i\big(\sum_{j=1}^{d} h_{ij}(x_j)\big)\): each \(h_{ij}:\mathbb{R}\!\to\!\mathbb{R}\) and \(g_i:\mathbb{R}\!\to\!\mathbb{R}\) is a small one-dimensional MLP; their inner/outer compositions are summed (optionally with a linear skip) to produce a scalar output.

\begin{table}[h]
\centering
\setlength{\tabcolsep}{8pt}
\renewcommand{\arraystretch}{1.2}
\begin{tabular}{|c|ccc|ccc|}
\hline
\rowcolor[HTML]{DAE8FC}
\cellcolor[HTML]{CBCEFB} &
\multicolumn{3}{c|}{\cellcolor[HTML]{DAE8FC}\textbf{Anchor (PINN)}} &
\multicolumn{3}{c|}{\cellcolor[HTML]{DAE8FC}\textbf{KAN}} \\
\cline{2-7}
\rowcolor[HTML]{FCE0DE}
\multirow{-2}{*}{\cellcolor[HTML]{CBCEFB}\textbf{European}} &
\multicolumn{1}{c|}{\cellcolor[HTML]{FCE0DE}\textbf{MAE}} &
\multicolumn{1}{c|}{\cellcolor[HTML]{FCE0DE}\textbf{RMSE}} &
\multicolumn{1}{c|}{\cellcolor[HTML]{FCE0DE}\textbf{EV}} &
\multicolumn{1}{c|}{\cellcolor[HTML]{FCE0DE}\textbf{MAE}} &
\multicolumn{1}{c|}{\cellcolor[HTML]{FCE0DE}\textbf{RMSE}} &
\multicolumn{1}{c|}{\cellcolor[HTML]{FCE0DE}\textbf{EV}} \\
\hline
Set 1 &
\multicolumn{1}{c|}{\textbf{4.5e-02}} &
\multicolumn{1}{c|}{\textbf{6.5e-02}} &
\multicolumn{1}{c|}{\textbf{9.3e-01}} &
\multicolumn{1}{c|}{8.1e-02} &
\multicolumn{1}{c|}{1.2e-01} &
\multicolumn{1}{c|}{5.7e-01} \\
\hline
Set 2 &
\multicolumn{1}{c|}{2.6e-01} &
\multicolumn{1}{c|}{4.4e-01} &
\multicolumn{1}{c|}{\textbf{9.3e-01}} &
\multicolumn{1}{c|}{\textbf{1.43e-01}} &
\multicolumn{1}{c|}{\textbf{1.7e-01}} &
\multicolumn{1}{c|}{6.1e-01} \\
\hline
Set 3 &
\multicolumn{1}{c|}{\textbf{8.5e-02}} &
\multicolumn{1}{c|}{\textbf{1.3e-01}} &
\multicolumn{1}{c|}{\textbf{9.8e-01}} &
\multicolumn{1}{c|}{3.4e-01} &
\multicolumn{1}{c|}{4.4e-01} &
\multicolumn{1}{c|}{6.2e-01} \\
\hline
Set 4 &
\multicolumn{1}{c|}{\textbf{2.3e-01}} &
\multicolumn{1}{c|}{\textbf{3.4e-01}} &
\multicolumn{1}{c|}{\textbf{9.6e-01}} &
\multicolumn{1}{c|}{1.2e+00} &
\multicolumn{1}{c|}{1.4e+00} &
\multicolumn{1}{c|}{8.4e-01} \\
\hline

\rowcolor[HTML]{DAE8FC}
\cellcolor[HTML]{CBCEFB} &
\multicolumn{3}{c|}{\cellcolor[HTML]{DAE8FC}\textbf{RNN}} &
\multicolumn{3}{c|}{\cellcolor[HTML]{DAE8FC}\textbf{ANN}} \\
\cline{2-7}
\rowcolor[HTML]{FCE0DE}
\multirow{-2}{*}{\cellcolor[HTML]{CBCEFB}} &
\multicolumn{1}{c|}{\cellcolor[HTML]{FCE0DE}\textbf{MAE}} &
\multicolumn{1}{c|}{\cellcolor[HTML]{FCE0DE}\textbf{RMSE}} &
\multicolumn{1}{c|}{\cellcolor[HTML]{FCE0DE}\textbf{EV}} &
\multicolumn{1}{c|}{\cellcolor[HTML]{FCE0DE}\textbf{MAE}} &
\multicolumn{1}{c|}{\cellcolor[HTML]{FCE0DE}\textbf{RMSE}} &
\multicolumn{1}{c|}{\cellcolor[HTML]{FCE0DE}\textbf{EV}} \\
\hline
Set 1 &
\multicolumn{1}{c|}{1.98e-01} &
\multicolumn{1}{c|}{2.73e-01} &
\multicolumn{1}{c|}{9.7e-01} &
\multicolumn{1}{c|}{2.5e-01} &
\multicolumn{1}{c|}{3.3e-01} &
\multicolumn{1}{c|}{6.8e-01} \\
\hline
Set 2 &
\multicolumn{1}{c|}{9.2e-01} &
\multicolumn{1}{c|}{8.4e-01} &
\multicolumn{1}{c|}{8.9e-01} &
\multicolumn{1}{c|}{9.6e-01} &
\multicolumn{1}{c|}{1.2e+00} &
\multicolumn{1}{c|}{8.8e-01} \\
\hline
Set 3 &
\multicolumn{1}{c|}{6.3e-01} &
\multicolumn{1}{c|}{8.4e-01} &
\multicolumn{1}{c|}{8.0e-01} &
\multicolumn{1}{c|}{7.9e-01} &
\multicolumn{1}{c|}{9.3e-01} &
\multicolumn{1}{c|}{8.5e-01} \\
\hline
Set 4 &
\multicolumn{1}{c|}{2.1e+00} &
\multicolumn{1}{c|}{2.8e+00} &
\multicolumn{1}{c|}{8.3e-01} &
\multicolumn{1}{c|}{2.6e+00} &
\multicolumn{1}{c|}{3.0e+00} &
\multicolumn{1}{c|}{7.6e-01} \\
\hline
\end{tabular}
\caption{Comparison of models for European call option pricing across four parameter sets of strike \(K\), volatility \(\sigma\), and time to maturity \(T\):
(i) \(K=30,\ \sigma=0.15,\ T=0.25\);
(ii) \(K=45,\ \sigma=0.20,\ T=0.5\);
(iii) \(K=45,\ \sigma=0.35,\ T=0.5\);
(iv) \(K=60,\ \sigma=0.30,\ T=1.0\).
Evaluation metrics: MAE, RMSE, and explained variance (EV).
}
\label{Tab:Euro_comp}
\end{table}


\begin{table}[h]
\centering
\setlength{\tabcolsep}{8pt}
\renewcommand{\arraystretch}{1.2}
\begin{tabular}{|c|ccc|ccc|}
\hline
\rowcolor[HTML]{DAE8FC}
\cellcolor[HTML]{CBCEFB} &
\multicolumn{3}{c|}{\cellcolor[HTML]{DAE8FC}\textbf{Anchor (PINN)}} &
\multicolumn{3}{c|}{\cellcolor[HTML]{DAE8FC}\textbf{KAN}} \\
\cline{2-7}
\rowcolor[HTML]{FCE0DE}
\multirow{-2}{*}{\cellcolor[HTML]{CBCEFB}\textbf{American}} &
\multicolumn{1}{c|}{\cellcolor[HTML]{FCE0DE}\textbf{MAE}} &
\multicolumn{1}{c|}{\cellcolor[HTML]{FCE0DE}\textbf{RMSE}} &
\multicolumn{1}{c|}{\cellcolor[HTML]{FCE0DE}\textbf{EV\_S}} &
\multicolumn{1}{c|}{\cellcolor[HTML]{FCE0DE}\textbf{MAE}} &
\multicolumn{1}{c|}{\cellcolor[HTML]{FCE0DE}\textbf{RMSE}} &
\multicolumn{1}{c|}{\cellcolor[HTML]{FCE0DE}\textbf{EV\_S}} \\
\hline
Set 1 &
\multicolumn{1}{c|}{\textbf{6.7e-02}} &
\multicolumn{1}{c|}{\textbf{1.2e-01}} &
\multicolumn{1}{c|}{\textbf{9.9e-01}} &
\multicolumn{1}{c|}{8.1e-02} &
\multicolumn{1}{c|}{1.2e-01} &
\multicolumn{1}{c|}{5.7e-01} \\
\hline
Set 2 &
\multicolumn{1}{c|}{\textbf{9.1e-02}} &
\multicolumn{1}{c|}{\textbf{1.7e-01}} &
\multicolumn{1}{c|}{9.3e-01} &
\multicolumn{1}{c|}{1.43e-01} &
\multicolumn{1}{c|}{1.7e-01} &
\multicolumn{1}{c|}{6.1e-01} \\
\hline
Set 3 &
\multicolumn{1}{c|}{\textbf{9.3e-02}} &
\multicolumn{1}{c|}{\textbf{1.3e-01}} &
\multicolumn{1}{c|}{\textbf{9.8e-01}} &
\multicolumn{1}{c|}{3.4e-01} &
\multicolumn{1}{c|}{4.4e-01} &
\multicolumn{1}{c|}{6.2e-01} \\
\hline
Set 4 &
\multicolumn{1}{c|}{\textbf{2.6e-01}} &
\multicolumn{1}{c|}{\textbf{3.4e-01}} &
\multicolumn{1}{c|}{\textbf{9.8e-01}} &
\multicolumn{1}{c|}{1.2e+00} &
\multicolumn{1}{c|}{1.4e+00} &
\multicolumn{1}{c|}{8.4e-01} \\
\hline

\rowcolor[HTML]{DAE8FC}
\cellcolor[HTML]{CBCEFB} &
\multicolumn{3}{c|}{\cellcolor[HTML]{DAE8FC}\textbf{RNN}} &
\multicolumn{3}{c|}{\cellcolor[HTML]{DAE8FC}\textbf{ANN}} \\
\cline{2-7}
\rowcolor[HTML]{FCE0DE}
\multirow{-2}{*}{\cellcolor[HTML]{CBCEFB}} &
\multicolumn{1}{c|}{\cellcolor[HTML]{FCE0DE}\textbf{MAE}} &
\multicolumn{1}{c|}{\cellcolor[HTML]{FCE0DE}\textbf{RMSE}} &
\multicolumn{1}{c|}{\cellcolor[HTML]{FCE0DE}\textbf{EV\_S}} &
\multicolumn{1}{c|}{\cellcolor[HTML]{FCE0DE}\textbf{MAE}} &
\multicolumn{1}{c|}{\cellcolor[HTML]{FCE0DE}\textbf{RMSE}} &
\multicolumn{1}{c|}{\cellcolor[HTML]{FCE0DE}\textbf{EV\_S}} \\
\hline
Set 1 &
\multicolumn{1}{c|}{1.1e-01} &
\multicolumn{1}{c|}{2.1e-01} &
\multicolumn{1}{c|}{9.8e-01} &
\multicolumn{1}{c|}{1.1e-01} &
\multicolumn{1}{c|}{2.0e-01} &
\multicolumn{1}{c|}{9.7e-01} \\
\hline
Set 2 &
\multicolumn{1}{c|}{7.6e-01} &
\multicolumn{1}{c|}{1.0e+00} &
\multicolumn{1}{c|}{8.8e-01} &
\multicolumn{1}{c|}{5.2e-01} &
\multicolumn{1}{c|}{7.1e-01} &
\multicolumn{1}{c|}{9.1e-01} \\
\hline
Set 3 &
\multicolumn{1}{c|}{1.05e+00} &
\multicolumn{1}{c|}{1.3e+00} &
\multicolumn{1}{c|}{8.5e-01} &
\multicolumn{1}{c|}{1.0e+00} &
\multicolumn{1}{c|}{1.3e+00} &
\multicolumn{1}{c|}{8.9e-01} \\
\hline
Set 4 &
\multicolumn{1}{c|}{3.30e+00} &
\multicolumn{1}{c|}{3.3e+00} &
\multicolumn{1}{c|}{7.5e-01} &
\multicolumn{1}{c|}{2.6e+00} &
\multicolumn{1}{c|}{3.0e+00} &
\multicolumn{1}{c|}{7.1e-01} \\
\hline
\end{tabular}
\caption{Comparison of models for American put option pricing across four parameter sets of strike \(K\), volatility \(\sigma\), and time to maturity \(T\):
(i) \(K=30,\ \sigma=0.15,\ T=0.25\);
(ii) \(K=45,\ \sigma=0.20,\ T=0.5\);
(iii) \(K=45,\ \sigma=0.35,\ T=0.5\);
(iv) \(K=60,\ \sigma=0.30,\ T=1.0\).
Evaluation metrics: MAE, RMSE, and explained variance (EV).
}
\label{Tab:Americ_comp}
\end{table}

Quantitative results (RMSE, MAE, EV) are summarized in Tables~\ref{Tab:Euro_comp} and \ref{Tab:Americ_comp}. Figure~\ref{fig:comparison_models} complements these with price–stock curves at \(t=0\) for both European (left) and American (right) cases: ground truth (pink), \textbf{KAN} (dashed dark green), \textbf{PINN-anchor} (dotted purple), \textbf{RNN} (cyan stars), and \textbf{ANN} (solid blue). Consistent with the tables, \textbf{PINN-anchor} and \textbf{KAN} generally outperform \textbf{ANN}/\textbf{RNN}, with visibly smaller deviations near the strike. Between \textbf{PINN-anchor} and \textbf{KAN}, accuracy is similar across most parameter sets; however, \textbf{KAN} trains markedly slower (about \(10\times\)) and exhibits sharper local variations—especially near the European strike—along with greater run-to-run variability, whereas \textbf{PINN-anchor} tends to train more stably.

\section{Conclusion}
Neural surrogates for option pricing are gaining traction \cite{dhiman2023physics, liu2023option, liu2024kolmogorov}, yet two gaps persist in the literature: (i) systematic comparisons to established numerical solvers, and (ii) uncertainty quantification (UQ) beyond point errors. We addressed both by formulating a physics-informed neural network (PINN) that solves the Black--Scholes (BS) PDE as a mesh-free, global surrogate over $(S,t)$ and by augmenting training with anchored ensembling to capture \emph{epistemic} uncertainty. To the best of our knowledge, this study is the first to provide confidence intervals for PINN solutions to option-pricing PDEs—an important contribution given that practitioners require not only point estimates but also well-calibrated confidence measures (in contrast, Monte Carlo methods routinely provide confidence bands).

Methodologically, our PINN approach departs from time-marching PDE solvers. Rather than stepping backward from maturity—where both error and uncertainty can accumulate—we learn a single solution over the entire $(t,S)$ domain. As a result, we do not observe a systematic increase in uncertainty at $t=0$; performance at $t=0$ is generally strong, particularly near the early-exercise boundary (the “kink”). Closer to maturity the solution develops sharper, less smooth features that neural networks find harder to learn; correspondingly, uncertainty is larger near the boundary at late times, yet absolute errors remain small (below \$0.30, i.e., $<\!5\%$ relative error). For American options, errors may grow with time, but our comparisons indicate that PINNs still outperform standard neural baselines (e.g., feed-forward and recurrent networks) on Black–Scholes solutions. Overall, these findings underline the suitability of PINNs for option pricing and show that our uncertainty bands closely track the error scale, yielding practically useful UQ.

Training a PINN is more computationally intensive than evaluating closed-form BS or running a single FD/MC solve; the benefit is amortization across queries and parameter sweeps. Performance depends on residual weighting and sampling (especially near $S\!\approx\!K$ and $t\!\to\!T$), and our UQ targets parameter (epistemic) uncertainty; modeling aleatoric noise would require an additional likelihood head. Future work includes (i) adaptive collocation or curricula to further reduce maturity-adjacent error, (ii) formal calibration studies (coverage with Wilson intervals, NLL, reliability) and conformal prediction for distribution-free intervals, (iii) higher-dimensional BS settings (baskets, stochastic rates), stochastic volatility and jump-diffusion dynamics, and (iv) broader benchmarks versus FD/PSOR/MC with runtime–accuracy trade-offs and ablations over anchor strength and ensemble size.



\bibliographystyle{unsrt}
{\footnotesize
\bibliography{PINN.bib}}



\end{document}